\documentclass[10pt,conference]{IEEEtran}

\newtheorem{thm}{Theorem}[section]
\usepackage{epsfig}
\usepackage{graphicx}
\usepackage{epstopdf}
\usepackage{amsmath}
\usepackage{amsfonts}
\usepackage{amssymb}
\usepackage{setspace}
\usepackage{lipsum}

\begin{document}

\title{Previous Messages Provide the Key to Achieve Shannon Capacity in a Wiretap Channel}
\author{\IEEEauthorblockN{Shahid Mehraj Shah}
\IEEEauthorblockA{Dept. of ECE, IISc Bangalore, India\\
Email: shahid@ece.iisc.ernet.in}
\and
\IEEEauthorblockN{Parameswaran S}
\IEEEauthorblockA{Dept. of ECE, IIT Kharagpur, India\\
Email: parameswaran.iitkgp@gmail.com}\and
\IEEEauthorblockN{Vinod Sharma}
\IEEEauthorblockA{Dept. of ECE, IISc Bangalore, India\\
Email: vinod@ece.iisc.ernet.in}}
\vspace{0.01cm}\maketitle
\begin{abstract}
We consider a wiretap channel and use previously transmitted messages to generate a secret key which increases the secrecy capacity. This can be bootstrapped to increase the secrecy capacity to the \textit{Shannon capacity} without using any feedback or extra channel while retaining the \textit{strong secrecy} of the wiretap channel.
\end{abstract}


\section{Introduction}
Shannon in 1948 in his seminal paper \cite{shannon1949} considered the problem of secure communication where he assumed that the legitimate receiver and  the eavesdropper receive the same information. Wyner \cite{wyner1975} assumed that the legitimate receiver and the eavesdropper receive different information due to channel differences and hence provided a coding scheme which achieves secrecy without using a key. In \cite{csizar1980} the authors studied the Broadcast channel with a secret message in a more general setting. 
The first work on secret key generation is reported in \cite{maurer_key}. In this paper the authors assume a public discussion channel for exchanging functions, and then to agree on a key. The eavesdropper ``hears" the whole conversation. \cite{ahlswede93} discusses two types of models: Source type model and Channel type model. Secret key generation with multiple terminals was studied  in \cite{csiszar2008}. 

Secret key generation via the sources and channels was investigated in \cite{diggavi} and\cite{prabhakaran}.

Wiretap channel with rate-distortion has been studied in \cite{yamamoto}. In \cite{ehsan} the authors have considered the wiretap channel with secure rate limited feedback. This feedback is used to agree on a secret key. Wiretap channel with shared key was studied in \cite{kang}.

Strong secrecy based secret key agreement was introduced in \cite{maurer_free}. For a detailed survey of Information theoretic security reader can refer to \cite{liang}. A Slow fading Wiretap channel with a secret key buffer was studied in \cite{helgamal}. The authors study the scenario where different secret messages are being transmitted in different slots and consider the equivocation of a message with only the outputs of the channel to the eavesdropper in the same slot. In \cite{helgamal_arx} the authors compute the equivocation of each message with the outputs of the channel to the eavesdropper in all these slots considered.

In this paper we study a model in which multiple messages are being transmitted by Alice in different slots. The equivocation of the messages in a slot is computed with all the channel outputs to the eavesdropper, as in \cite{helgamal_arx}. In the first slot Alice transmits a codeword using wiretap coding, as in \cite{wyner1975}. Only Bob can decode this message but not the eavesdropper. Thus, in the next slot, we can use this message as a key to transmit the next message and also use wiretap coding. This increases the secret message rate to twice the secrecy capacity of the wire-tap channel. This whole message can be used as a key for the next slot. This we repeat till we achieve the secret key rate equal to the capacity of the main channel, and then the rest of the communication takes place using secret key at that rate. We will show that this does not increase the information leakage rate to eve.

The rest of the paper is organised as follows: Channel model and the problem statement are provided in Section II. In Section III we provide our coding and decoding scheme and show that it can provide Shannon capacity without sacrificing secrecy. In Section IV we apply our coding scheme on a Gaussian wiretap channel. Section V concludes this paper.

A note about the notation: capital letters, like $W$ will denote a random variable and the corresponding small letter  $w$ its realization. An $n$-length vector $(A_1, A_2,\ldots, A_n)$will be denoted as $A^n $. Information theoretic notation will be same as in  \cite{elgamal}.
\section{Channel Model And Problem Statement}
\setlength{\jot}{6pt}
We consider a discrete time, memoryless, degraded wiretap channel, where Alice wants to transmit messages to Bob. There is an eavesdropper (Eve) who is passively ``listening"(Fig. 1). We want to keep Eve ignorant of the messages. 

\begin{figure}
\epsfig{figure=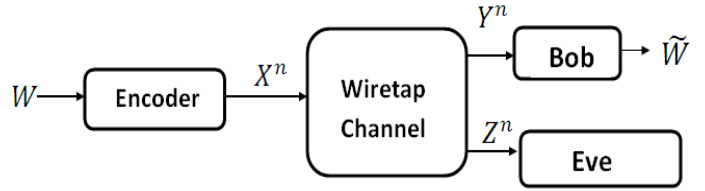,height=2.5cm,width=9cm}
\caption{The Wiretap channel}
\label{fig1}
\end{figure}

Formally, Alice wants to communicate messages $W \in \mathcal{W}=\{1,2,\ldots,2^{nR_s}\}$ reliably over the Wiretap channel to Bob, while ensuring that Eve is not able to decode them. Here $R_s$ is the secrecy capacity of the wiretap channel defined as 
\begin{equation}
R_s=\max_{p(x)}\left[I(X;Y)-I(X;Z)\right].
\end{equation}
We assume $R_s>0$. The transition probability matrix of the channel is $p(y,z|x)$. At time $i$, $X_i$ is the channel input and the legitimate receiver (Bob) and Eve receive the channel outputs  $Y_i$ and $Z_i$ respectively, where $X_i \in \mathcal{X}, Y_i \in \mathcal{Y}, Z_i \in \mathcal{Z}$. The messages $W_k$ are generated uniformly from $\mathcal{W}$ and $\{W_m, m\geq 1\}$ is an independent sequence. One or more message is encoded into an $n$ length codeword. A mini-slot consists of $n$ channel uses. In our scheme, the first slot consists of only one mini-slot. Then upto $\lambda$ slots, each slot consists of 2 mini-slots where 
\begin{equation}
\lambda \triangleq \left[ \frac{C}{R_s} \right] ,
\end{equation}
and $C$ is the capacity of Alice-Bob channel and $[x]$ is the integer part of $x$. For simplicity, we take $\frac{C}{R_s}$ as integer.
Finally, after $\lambda$ slots each slot has only one mini-slot. The message $\overline{W_k}$ to be transmitted in slot $k$ consists of one or more messages $W_m$. The codeword for message $\overline{W}_k$ (for $1<k\leq \lambda$) is denoted by $X_k^{2n}=\{X_{k1},\ldots,X_{2kn}\}$ or $X_k^n$ depending on the length of the slot. To increase the secrecy rate, the transmitter uses the secret message $\overline{W}_k$ transmitted in slot $k$ as the key for transmitting the message in slot $k+1$. 

\subsection{Encoder:} To transmit message $\overline{W}_{k+1}$ in slot $k+1$, the encoder has two parts
\begin{equation}
f_s:\mathcal{W}\rightarrow \mathcal{X}^{n}, f_d:\mathcal{W} \times \mathcal{K} \rightarrow \mathcal{X}^{n},
\end{equation}
where $X \in \mathcal{X}$, and $\mathcal{K}$ is the set of secret keys generated and $f_s$ is the Wiretap encoder, as in \cite{wyner1975}.
For $f_d$ one can use various encoders studied for transmission with secret key. We use the following: Take binary version of the message and $XOR$ with the binary version of the key. Encode the resulting encrypted message with an optimal usual channel encoder.

In the first slot, a message is encoded using the wiretap code only. From second slot onwards (till slot $\lambda$), both wiretap encoder $f_s$ and deterministic encoder $f_d$ are used. Thus in slot $k$, we can say that $k$ messages from $\mathcal{W}$ are sent, 1 using wiretap coding and $k-1$ using a key of rate $(k-1)R_s$. Of course the overall coding rate should not exceed capacity $C$ of the main channel (Alice $\rightarrow$ Bob). After slot $\lambda$ only one mini-slot is used with a key of rate $C$ (assuming $C$ is multiple of $R_s$, otherwise the key rate will be $\left[\frac{C}{R_s}\right]R_s$).

\subsection*{Decoder} For the first slot of communication, the decoder function at Bob is 
\begin{equation}
\phi_1 : \mathcal{Y}^{2n} \rightarrow \mathcal{W}.
\end{equation}
From second slot onwards, the decoder also has a secret key (which is generated in the previous slot). Thus, the decoder is 

\begin{equation}
\phi_i : \mathcal{Y}^n \times \mathcal{K} \rightarrow \mathcal{W}^j 
\end{equation}
for time slot $i$, with $j=min(i,\frac{C}{R_s})$.
The probability of error for this code is:
\begin{equation}
P_e^{(n)} = Pr\{\widehat{W}\neq \overline{W}\}
\end{equation}
where $\widehat{W}$ is the decoded message.

\textit{Leakage rate} is $R_L^n = \frac{1}{n}I(\overline{W};Z^{2n})$. This is the rate at which information is getting leaked to Eve.

\textit{Definition 1}: A Leakage-rate pair $(R_L,R)$ is said to be achievable if there exists a sequence of $(2^{nR},n)$-codes such that $P_e^{(n)} \rightarrow 0$ and $\limsup_{n\rightarrow \infty}R_L^n \leq R_L$ as $n \rightarrow \infty$. Actually in slot $k\geq 2$ we will consider the leakage rate $\frac{1}{2n}I(\overline{W}_m;Z_1^n,Z_2^{2n},\ldots,Z_k^{2n})$.

We will be concerned about the rate achievable when $R_L=0$.
\section{Capacity of Wiretap Channel}

\begin{thm}
The rate $(0,C)$ is achievable  for all slots $k\geq\lambda$.
\end{thm}
\textit{Proof of Achievability}: In the first slot of communication, Alice picks  message $W_1$ from $\mathcal{W}$ and transmits this message using $(n,2^{nR_s})$-Code. Bob decodes this message as $\widehat{W}_1$.

In the second slot using the previous message, $\overline{W}_1=W_1$, as a key (with key rate $R_k=R_s$) Alice transmits message $\overline{W}_2=(W_{21},W_{22})$, where $W_{21}=W_{2}, W_{22}=W_3$ are taken from the $iid$ sequence $\{W_k,k\geq 1\}$. To transmit this message, we use the following coding strategy:
\\
The first message $W_{21}$ is encoded to $X_{21}^{n}$ using wiretap code. The second message $W_{22}$ is first encrypted to produce the cipher using one-time pad with the previous message as secret key, i.e., $K=W_1$ and the cipher is $\widetilde{W}_{22}=W_{22}\bigoplus W_1$. We encode this encrypted message to $X_{22}^{n}$ using a point-to-point optimal channel code, to transmit it over the channel (practically, one can use LDPC or Turbo Codes).
Hence the overall codeword that is transmitted over the channel is $X_{21}^{n}X_{22}^{n}=X_2^{2n}$
with the overall rate $R_s$. 

In slot 3, $\overline{W}_3=(W_{31},W_{32})$ is transmitted where $W_{31}=W_4$ and $W_{32}$ is $(W_5,W_6)$, i.e., $W_{32}$ consists of two messages from $\mathcal{W}$. $W_{31}$ is encoded as $X_{31}^n$ using wiretap coding. $W_{32}$ is encoded via the key $\overline{W}_2$: Using usual optimal channel code at rate $2R_s$, encode $\overline{W}_2\bigoplus W_{32}$, provided of course $2R_s<C$.

We continue this till $\lambda -1$ slots. In slot $\lambda-1$, we transmit message $(W_{\lambda -1,1}, W_{\lambda -1,2},\ldots,W_{\lambda -1,\lambda -1})$. We will use the previous message $(W_{\lambda -2,2},\ldots, W_{\lambda -2,\lambda -2})$ as the key with the key rate $R_k = (\lambda -1)R_s$. Message $W_{\lambda-1,1}$ is sent via wiretap coding and the rest via the secret key.  Now we achieve the total rate, 
 
 \begin{equation}
 \frac{1}{2}\left(R_s + (\lambda -1)R_s\right) = \frac{1}{2}\left(R_s+C\right).
 \end{equation}
In the next slot we will only have $n$ channel uses and use only the key with rate $C$ and no wiretap coding. This provides us the secret rate of $C$. From then onward we repeat this codebook with the key as the previous message and obtain a secrecy rate of $C$.

Bob decodes the message as follows. In slot $k$, (for $1<k<\lambda$) $Y_{k1}^n$ is decoded via usual wiretap decoding while $Y_{k2}^n$ is decoded first by the channel decoder and then $XOR$ed with $\widehat{W}_{k-1}$. The probability of error for Bob goes to zero as $n\rightarrow \infty$. There is a small issue of error propagation due to using the previous message as key: Let $\epsilon_n$ be the message error probability for the wiretap encoder and let $\delta_n$ be the message error probability due to the channel encoder for $W_{k}$. Then $\epsilon_n \rightarrow 0$ and $\delta_n \rightarrow 0$ as $n\rightarrow \infty$. For the $k^{th}$ slot, $1<k<\lambda-1$, we have $P(\overline{W}_k\neq\widehat{W}_k)\leq Pr($Error in decoding $ W_{k1})+ Pr($Error in decoding $ \widetilde{W}_{k2})+Pr($Error in decoding $ \overline{W}_{k-1}) \leq k\epsilon_n + (k-1)\delta_n$. Thus the error increases with $k$. But restarting (as in slot 1) after some $k$ slots (somewhat large compared to $\lambda$) as in slot 1 will ensure that $P(\overline{W}_k\neq\widehat{W}_k)\rightarrow 0$ as $n\rightarrow \infty$.

Next we compute the leakage rate for Eve. In slot 1, wire-tap coding is used. Therefore, 
$\frac{1}{n}I(\overline{W}_1;Z_1^n)\rightarrow 0$, as $n\rightarrow \infty$.
In the following we fix an $\epsilon>0$ and take $n$ such that $I(\overline{W}_1;Z_1^n)\leq n\epsilon$.

In slot 2 we want to show that 
\begin{equation}
\frac{1}{n}I(\overline{W}_1;Z_1^n,Z_2^{2n})\rightarrow 0
\end{equation}
and
\begin{equation}
\frac{1}{n}I(\overline{W}_2;Z_1^n,Z_2^{2n})\rightarrow 0,
\end{equation}
 as $n\rightarrow \infty$. We have, 
\begin{equation}
I(\overline{W}_1;Z_1^n,Z_2^{2n})=I(\overline{W}_1;Z_1^n)+I(\overline{W}_1;Z_2^{2n}|Z_1^n).
\end{equation}
Since $\overline{W}_2=(W_{21},W_{22}\bigoplus \overline{W}_1)$ $\perp$ $\overline{W}_1$, $Z_2^{2n}\perp (\overline{W}_1,Z_1^n)$ ($X\perp Y$ will denote $X$ is independent of $Y$). Therefore,
\begin{equation}
\label{eq3}
I(\overline{W}_1;Z_2^{2n}|Z_1^n)=0.
\end{equation}
Also, $I(\overline{W}_1;Z_1^n) \leq n\epsilon$ and hence
\begin{equation}
\label{eq4}
I(\overline{W}_1;Z_1^n,Z_2^{2n})\leq n\epsilon.
\end{equation} 
 
Next consider
\begin{flalign}
&I(\overline{W}_2;Z_1^n,Z_2^{2n}) \nonumber &\\
&=I(\overline{W}_2;Z_1^n)+I(\overline{W}_2;Z_2^{2n}|Z_1^n) \label{eq1}. &
\end{flalign}

Since $\overline{W}_2 \perp Z_1^n$, $I(\overline{W}_2;Z_1^n)=0$. Now consider the second term in (\ref{eq1}), 
\begin{flalign}
&I(\overline{W}_2;Z_2^{2n}|Z_1^n)=I(W_{21},W_{22};Z_2^{2n}|Z_1^n)  \nonumber &\\
&=I(W_{21};Z_2^{2n}|Z_1^n)+I(W_{22};Z_2^{2n}|Z_1^n,W_{21}). \label{eq1.3}
\end{flalign}
Also,
\begin{flalign}
&I(W_{21};Z_2^{2n}|Z_1^n) \nonumber &\\
&=I(W_{21};Z_{22}^n|Z_1^n)+I(W_{21};Z_{21}^{n}|Z_1^n,Z_{22}^n). \label{eq1.4}
\end{flalign}
Since $W_{21} \perp (Z_{22}^n,Z_1^n)$, 
\begin{equation}
I(W_{21};Z_{22}^n|Z_1^n)=0.  \label{eq1.5}
\end{equation}
Furthermore, $(W_{21},Z_{21}^n)\perp(Z_1^n,Z_{22}^n)$, implies
\begin{equation}
I(W_{21};Z_{21}^{n}|Z_1^n,Z_{22}^n)=I(W_{21};Z_{21}^n)\leq n\epsilon.  \label{eq1.6}
\end{equation}

From (\ref{eq1.4}), (\ref{eq1.5}) and (\ref{eq1.6})
\begin{equation}
I(W_{21};Z_{22}^{2n}|Z_1^n)\leq n\epsilon.  \label{eq1.7}
\end{equation}
Now we consider second term of (\ref{eq1.3}),
\begin{align}
&I(W_{22};Z_2^{2n}|Z_1^n,W_{21})=I(W_{22};Z_{21}^{n},Z_{22}^n|Z_1^n,W_{21}) \nonumber &\\
&=I(W_{22};Z_{21}^{n}|Z_1^n,W_{21})+I(W_{22};Z_{22}^n|Z_1^n,Z_{21}^{n},W_{21}). \label{eq1.8}
\end{align}
Since $W_{22}\perp(Z_{21}^n,Z_1^n,W_{21})$, 
\begin{equation*}
I(W_{22};Z_{21}^n|Z_1^n,W_{21})=0.
\end{equation*}
Also $(Z_{21}^n,W_{21})\perp(W_{22},Z_{22}^n,Z_1^n)$, implies,
\begin{equation}
I(W_{22};Z_{22}^n|Z_{21}^n,Z_1^n,W_{21})=I(W_{22};Z_{22}^n|Z_1^n). 
\end{equation}
But $W_{22}\perp Z_1^n$ implies
\begin{flalign}
&I(W_{22};Z_{22}^n|Z_1^n)=I(W_{22};Z_1^n,Z_{22}^n) \nonumber &\\
&~~~~~~~~~~~~~~~~~~~=I(W_{22};Z_{22}^n)+I(W_{22};Z_1^n|Z_{22}^n)\nonumber &\\
&~~~~~~~~~~~~~~~~~~~=I(W_{22};Z_1^n|Z_{22}^n),
\end{flalign}
because $I(W_{22};Z_{22}^n)=0$. Now observe that the following Markov relationship holds
\begin{equation}
Z_1^n\longleftrightarrow W_1 \longleftrightarrow (W_1,W_{22}) \longleftrightarrow Z_{22}^n. \label{markov} 
\end{equation}
Therefore,
\begin{flalign}
&I(Z_1^n;W_{22}|Z_{22}^n)\leq I(Z_1^n;W_{22},W_1|Z_{22}^n)\leq I(Z_1^n;W_{22},W_1) \nonumber &\\
&~~~~~~~~~~~~~~~~~~~=I(Z_1^n;W_1)+I(Z_1^n;W_{22}|W_1).
\end{flalign}
Because of wiretap coding $I(Z_1^n;W_1)\leq n\epsilon$. Also from (\ref{markov}),
\begin{equation*}
I(Z_1^n;W_{22}|W_1)=0.
\end{equation*}
Hence,
\begin{equation}
I(Z_1^n;W_{22}|Z_{22}^n) \leq n\epsilon.
\end{equation}

Along with (\ref{eq1.7}), this implies that $I(\overline{W}_2;Z_1^n,Z_2^{2n})\leq 2n\epsilon$.

Next we use mathematical induction to show that $\frac{1}{n}I(\overline{W}_m;Z_1^n,Z_2^{2n},\ldots,Z_{k+1}^{2n})\rightarrow 0$ for all $m\leq k+1, k\geq 1$. We use the notation,
\begin{equation}
Z^{(m)}=(Z_1^n,Z_2^{2n},\ldots,Z_m^{2n}),~~~~~ m=1,2,\ldots
\end{equation}

We show
\begin{equation}
\frac{1}{n}I(\overline{W}_m;Z^{(k+1)})\leq 2\epsilon,
\end{equation}
for $m=1,\ldots k+1$
given,
\begin{equation}
\frac{1}{n}I(\overline{W}_m;Z^{(k)})\leq 2\epsilon, \label{ind1}
\end{equation}
for $m=1,\ldots,k$.

For $m=1,\ldots k$,

\begin{equation}
I(\overline{W}_m;Z^{(k+1)})=I(\overline{W}_m;Z^{(k)})+I(\overline{W}_m;Z_{k+1}^{2n}|Z^{(k)}). \label{ind2}
\end{equation}
From (\ref{ind1}) $I(\overline{W}_m;Z^{(k)}) \leq n\epsilon$.

The second term,
\begin{align}
&I(\overline{W}_m;Z_{k+1}^{2n}|Z^{(k)}) \nonumber &\\
&~~=I(\overline{W}_m;Z_{k+1,1}^n,Z_{k+1,2}^n|Z^{(k)}) \nonumber &\\
&~~=I(\overline{W}_m;Z_{k+1,1}^n|Z^{(k)})+I(\overline{W}_m;Z_{k+1,2}^n|Z^{(k)},Z_{k+1,1}^n). \label{new1}
\end{align}
Also,
\begin{flalign}
&I(\overline{W}_m;Z_{k+1,1}^n|Z^{(k)})=I(W_{m1},W_{m2};Z_{k+1,1}^n|Z^{(k)}) \nonumber &\\
&~~~~~~~~~~~~~~~~~~~~~~~~=I(W_{m1};Z_{k+1,1}^n|Z^{(k)}) \nonumber &\\
&~~~~~~~~~~~~~~~~~~~~~~~~+I(W_{m2};Z_{k+1,1}^n|Z^{(k)},W_{m1}). \label{new2}
\end{flalign}
Now since $(W_{m1},Z_{k+1,1})\perp Z^{(k)}$,
\begin{equation}
I(W_{m1};Z_{k+1,1}^n|Z^{(k)})=I(W_{m1};Z_{k+1,1})=0.  \label{new3}
\end{equation}

~~~Next consider the second term of (\ref{new2}), $I(W_{m2};Z_{k+1,1}^n|Z^{(k)},W_{m1})$. Since $Z_{k+1,1}^n\perp (Z^{(k)},\overline{W}_m)$, 
\begin{equation}
I(W_{m2};Z_{k+1,1}^n|Z^{(k)},W_{m1})=0. \label{new4}
\end{equation}
Hence from (\ref{new2}), (\ref{new3}) and (\ref{new4}), we get
\begin{equation}
I(\overline{W}_{m};Z_{k+1,1}^n|Z^{(k)})=0. \label{new5}
\end{equation}
Now we consider 
\begin{equation*}
I(\overline{W}_{m};Z_{k+1,2}^n|Z^{(k)},Z_{k+1,1}^n),~~~ m=1,\ldots,k.
\end{equation*}
When $m=k$ the following Markov relation holds,

\begin{align}
&(Z^{(k)},Z_{k+1,1}^n)\leftrightarrow (W_{k1},W_{k2}) \nonumber &\\
&~~~~~~~~~~~~~~~~~\leftrightarrow (W_{k1},W_{k2},W_{k+1,2})\leftrightarrow Z_{k+1,2}^n. \label{markov_2}
\end{align}
Thus, by Markov inequality,
\begin{flalign}
&I(\overline{W}_{k};Z_{k+1,2}^n|Z^{(k)},Z_{k+1,1}^n)\leq I(W_{k1},W_{k2};Z_{k+1,2}^n) \nonumber &\\
&~~~~~~~~~~~~~~~~~~~~~~~~~~~~~~~~~=I(W_{k2};Z_{k+1,2}^n)=0.
\end{flalign}

Therefore from (\ref{ind2}), we get $I(\overline{W}_k;Z^{(k+1)})\leq n\epsilon$. 
Now for $m<k$, since $Z^{(m-1)}\perp(\overline{W}_m,Z_{k+1}^{2n},Z_m^{2n},\ldots,$ $Z_{k}^{2n})$, we have
\begin{flalign}
&I(\overline{W}_{m};Z_{k+1,2}^n|Z^{(k)},Z_{k+1,1}^n) \nonumber &\\
&~=I(\overline{W}_{m};Z_{k+1,2}^n|Z_m^{2n},\ldots Z_{k}^{2n},Z_{k+1,1}^n).
\end{flalign}
From the following Markov relation
\begin{equation}
(Z_m^{2n},\ldots,Z_k^{2n},Z_{k+1,1}^n)\leftrightarrow (\overline{W}_{m}, \overline{W}_{m+1},...,\overline{W}_{k+1})\leftrightarrow Z_{k+1,2}^n,    \label{markov_3}
\end{equation}
we get
\begin{flalign}
&I(\overline{W}_{m};Z_{k+1,2}^n|Z_m^{2n},\ldots,Z_k^{2n},Z_{k+1,1}^n)\nonumber &\\
&~~\leq I(\overline{W}_{m},\overline{W}_{m+1},\ldots,\overline{W}_{k+1};Z_{k+1,2}^n|Z_m^{2n},\ldots,Z_k^{2n},Z_{k+1,1}^2) \nonumber &\\
&~~\leq I(\overline{W}_{m},\overline{W}_{m+1},\ldots,\overline{W}_{k};Z_{k+1,2}^n)=0.  \label{new_7}
\end{flalign}
Thus we obtain 
\begin{equation}
I(\overline{W}_m;Z^{(k+1)})\leq n\epsilon, \label{cite_end}
\end{equation}
for $m=1,\ldots,k.$


Now consider
\begin{align}
&I(\overline{W}_{k+1};Z^{(k+1)})=I(W_{k+1,1},W_{k+1,2};Z^{(k+1)}) \nonumber &\\
&~=I(W_{k+1,1};Z^{(k+1)})+I(W_{k+1,2};Z^{(k+1)}|W_{k+1,1}). \label{final1}
\end{align}

We consider the first term of (\ref{final1}),
\begin{flalign}
&I(W_{k+1,1};Z^{(k+1)})\nonumber &\\
&~=I(W_{k+1,1};Z^{(k)})+I(W_{k+1,1};Z_{k+1}^{2n}|Z^{(k)}).  \label{final2}
\end{flalign}
Since $W_{k+1,1}\perp (Z_1^n,\ldots,Z_k^{2n})$,
\begin{equation}
I(W_{k+1,1};Z^{(k)})=0.  \label{final4}
\end{equation}
Also,
\begin{flalign}
&I(W_{k+1,1};Z_{k+1}^{2n}|Z^{(k)}) \nonumber &\\ 
&~~=I(W_{k+1,1};Z_{k+1,2}^n|Z^{(k)})+I(W_{k+1,1};Z_{k+1,1}^n|Z^{(k)},Z_{k+1,2}^n)\nonumber &\\
&~~=0+n\epsilon.  \label{final3}
\end{flalign}

Thus,
\begin{equation}
I(W_{k+1,1};Z^{(k+1)})\leq n\epsilon.  \label{cite_end_1}
\end{equation}

Now we consider second term in (\ref{final1}),
\begin{flalign}
&I(W_{k+1,2};Z^{(k+1)}|W_{k+1,1}) \nonumber &\\
&~~=I(W_{k+1,2};Z^{(k)},Z_{k+1,1}^n,Z_{k+1,2}^n|W_{k+1,1})\nonumber &\\
&~~=I(W_{k+1,2};Z_{k+1,1}^n|W_{k+1,1}) \nonumber &\\
&~~+I(W_{k+1,2};Z^{(k)},Z_{k+1,2}^n|W_{k+1,1},Z_{k+1,1}^n).\label{final5}
\end{flalign}

Since $W_{k+1,2}\perp(W_{k+1,1},Z_{k+1,1}^n)$,
\begin{equation*}
I(W_{k+1,2};Z_{k+1,1}^n|W_{k+1,1})=0.
\end{equation*}
Also we note that $(W_{k+1,1},Z_{k+1,1}^n)\perp(W_{k+1,2},Z^{(k)},Z_{k+1,2}^n)$ and $W_{k+1,2}\perp Z^{(k)}$ and hence (\ref{final5}) becomes
\begin{flalign}
&I(W_{k+1,2};Z^{(k)},Z_{k+1,2}^n|W_{k+1,1},Z_{k+1,1}^n) \nonumber &\\
&~~=I(W_{k+1,2};Z^{(k)},Z_{k+1,2}^n) \nonumber &\\
&~~=I(W_{k+1,2};Z^{(k)})+I(W_{k+1,2};Z_{k+1,2}^n|Z^{(k)}) \nonumber &\\
&~~=I(W_{k+1,2};Z_{k+1,2}^n|Z^{(k)}). \label{final5.1}
\end{flalign}
Also since $Z^{(k-1)}\perp (W_{k+1,2},Z_{k+1,2}^n,Z_k^{2n})$, 
\begin{equation}
I(W_{k+1,2};Z_{k+1,2}^n|Z^{(k)})=I(W_{k+1,2};Z_{k+1,2}^n|Z_k^{2n}). \label{final5.2}
\end{equation}
But $W_{k+1,2}\perp Z_k^{2n}$ implies
\begin{flalign}
&I(W_{k+1,2};Z_{k+1,2}^n|Z_k^{2n})=I(W_{k+1,2};Z_{k+1,2}^n,Z_k^{2n})\nonumber &\\
&~~~=I(W_{k+1,2};Z_{k+1,2}^n)+I(W_{k+1,2};Z_k^{2n}|Z_{k+1,2}^n) \nonumber &\\
&~~~=I(W_{k+1,2};Z_k^{2n}|Z_{k+1,2}^n).
\end{flalign}
Now note that the following Markov relationship holds
\begin{equation}
Z_k^{2n}\longleftrightarrow \overline{W}_k \longleftrightarrow (\overline{W}_k,\overline{W}_{k+1,2})\longleftrightarrow Z_{k+1,2}^{2n}  \label{markov2}
\end{equation}
and also $W_{k+1,2}\perp(\overline{W}_k,Z_k^{2n})$. Therefore,
\begin{flalign}
&I(W_{k+1,2};Z_k^{2n}|Z_{k+1,2}^n)\leq I(W_{k+1,2}\overline{W}_k;Z_k^{2n}|Z_{k+1,2}^n) \nonumber &\\
&~~~\leq I(W_{k+1,2},\overline{W}_k;Z_k^{2n})\nonumber &\\
&~~~\leq I(\overline{W}_{k};Z_k^{2n})+I(W_{k+1,2};Z_k^{2n}|\overline{W}_k)\leq n\epsilon +0.  \label{final6}
\end{flalign}
From (\ref{final1}) and (\ref{cite_end_1}), now we obtain
\begin{equation}
I(\overline{W}_{k+1};Z^{(k+1)})\leq 2n\epsilon.~~~~~~~~~~~~~~~~~~~~~~\square
\end{equation}

\textit{Comment}: We can obtain Shannon capacity even with \textit{strong secrecy}. To do that we have to use \textit{information reconciliation} and \textit{privacy amplification} in the first slot after transmitting message $\overline{W}_1$ using wiretap coding, as is done in \cite{maurer_free} and \cite{bloch_2008}. In the subsequent blocks we use both the wiretap encoder and the deterministic encoder. Wiretap encoder is used to transmit one message using wiretap coding and the deterministic encoder is used for transmitting the other message (which is encrypted with the secret key generated in strong secure sense in the previous slot) using usual channel codes. Here also we need to modify the wiretap encoder by incorporating information reconciliation and privacy amplification for the message which we transmit using wiretap code. In this way we ensure that in every slot we generate the secret key for the next slot which is strongly secure, i.e, in $k^{th}$ slot, we have
\begin{equation}
\label{eq24}
I(\overline{W}_m;Z_1^n,\ldots,Z_k^{2n}) \rightarrow 0, m=1,\ldots,k.
\end{equation}

Proof of (\ref{eq24}) follows as in Theorem 3.1
\section{Examples}
\subsection{Gaussian Wiretap Channel}
Consider Additive White Gaussian Noise Channel (AWGN) wiretap channel with average power constraint $P$. The noise variance at Bob and Eve are $\sigma_b^2$ and $\sigma_e^2$ respectively, with $\sigma_b^2<\sigma_e^2$. The channel codes are chosen from Gaussian codebooks as $X \thicksim \mathit{N}(0,P) $. Then, from \cite{hellman1976}

\begin{equation}
\label{eq25}
R_{s} = \frac{1}{2}\mathsf{log} \left(1+\frac{P}{\sigma_b^2} \right) - \frac{1}{2}\mathsf{log} \left(1+\frac{P}{\sigma_e^2} \right).
\end{equation}
The key rate $R_K$ in slot 2 is $R_s$, in slot 3 is $2R_s$ and so on. After slot $\lambda$, where
\begin{equation}
\label{eq26}
\lambda=\frac{\frac{1}{2}\log\left(1+\frac{P}{\sigma_b^2}\right)}{\frac{1}{2}\mathsf{log} \left(1+\frac{P}{\sigma_b^2} \right) - \frac{1}{2}\mathsf{log} \left(1+\frac{P}{\sigma_e^2} \right)},
\end{equation}
the capacity will reach $\frac{1}{2}\log\left(1+\frac{P}{\sigma_b^2}\right)$ provided $\frac{C}{R_s}$ is integer valued.

\section{Acknowledgement}
The authors would like to thank Prof. Vinod M. Prabhakaran (TIFR Mumbai) for his valuable comments.
\section{Conclusion}
In this paper we have achieved secrecy rate equal to the main channel capacity by using previous secret messages as key for transmitting the current message. This can be done while still retaining \textit{strong secrecy}.

\end{document}